\documentclass[preprint]{revtex4}
\usepackage[latin9]{luainputenc}
\usepackage{amsmath}
\usepackage{graphicx}
\usepackage{amsfonts}
\usepackage{amsfonts}

\makeatletter
\expandafter\let\csname equation*\endcsname\relax
\expandafter\let\csname endequation*\endcsname\relax
\makeatother

\begin{document}

\title{Macroscopic approach to N-qudit systems}
\author{C. Muñoz, I. Sainz, A.B. Klimov}

\begin{abstract}
We develop a general scheme for an analysis of macroscopic qudit systems: a)
introduce a set of collective observables, which characterizes the
macroscopic properties of qudits in an optimal way; b) construct projected $%
\tilde{Q}$-functions for $N$ qudit systems, containing full macroscopic
information; c) propose a collective tomographic protocol both for a general
and symmetric $N$-qudit states. The example of $N$-qutrit is analyzed in
details and compared to $N$-qubit case.
\end{abstract}

\maketitle

\address{ Dept. de Fisica, Universidad de Guadalajara, 44420 Guadalajara,
Mexico} 
\email{klimov@cencar.udg.mx}



\section{Introduction}

Basic problems arising in the analysis of macroscopic quantum systems: a)
optimization of tomographic schemes in order to reduce the number of
measurements \cite{CS}, \cite{TE}, \cite{MUB}; b) processing of available
data so that a meaningful information about the state of the system can be
efficiently extracted \cite{Standard}. Both of these tasks are rather
complicated since the number of parameters (the number of measurement
outcomes) required for a complete characterization of a multipartite quantum
state grows exponentially with the number of particles. An intuitively
appealing attempt to employ the statistical description of $N$-qudit ($d$%
-level) system, by mapping qudits states into distributions in a
finite-dimensional grid \cite{Wootters87}, \cite{simplect},\cite{DCS},
results inconvenient in the large $N$ limit due to an overwhelming
complexity of the resulting discrete functions. Even in the simplest case of
$N$-qubit system, the distributions corresponding to relatively simple
quantum states have rather complicated structure (usuall represented in form
of randomly located peaks \cite{WFpic}). In addition, the distributions in
discrete phase-space do not have natural macroscopic limits, i.e. they do
not acquire smooth shapes in the limit $N\gg1$, which makes very difficult
to study their analytical porperties.

While $\sim d^{2N}$\ parameters are needed for a full microscopic
description of a generic $N$ qudit state, the \textit{global} properties can
be captured by a significantly smaller number of collective observables.
These observables are invariant under particle permutations and thus provide
only partial information about the system. On the other hand, frequently
only this type of \textit{symmetric} correlation functions can be
efficiently assessed (e.g. due to particle indistinguishability) in \
macroscopic quantum systems \cite{bose}.

A quantum state characterization is strongly connected to the specific set
of the measurable collective variables. In $N$ qubit systems, the moments of
the collective spin operators
\begin{equation}
S_{x,y,z}=\sum_{i=1}^{N}\sigma_{x,y,z}^{(i)},  \label{S}
\end{equation}
provide the global information, allowing to access all $SU(2)$ invariant
subspaces appearing in the tensor decomposition $\mathcal{H}_{2}^{\otimes N}$
(without distinguishing subspaces of the same dimension). The results of
such collective measurements can used for density matrix estimation \cite%
{colTom}.

From an analytical perspective, the macroscopic features of an $N$ qubit
system are completely described by the so-called projected $\tilde{Q}$
-function, defined in a three-dimensional discrete space. This $\tilde{Q}$
-function contains full and non-redundant information about results of any
collective measurement in an arbitrary (not necessarily symmetric) state
\cite{macros}, \cite{galo}, and allows to outline collective kinematic and
dynamic properties in the limit of large number of qubits.

The situation is more complicated for macroscopic systems containing a large
number of qudits. Although qudit systems seem to be useful in quantum
information protocols \cite{qdits}, only a limited number of experimental
reconstructions of one- and bi-partite qudit systems were reported \cite%
{1qditTom} - \cite{quditTom15}. Extension to multipartite qudit systems is
quite challenging due to a rapid growth of the number of required
experimental setups \cite{qditTom}, and the deficiency of appropriate
theoretical tools to deal with $N\gg1$ qudit systems. \ In addition, in this
case (in contrast to qubit systems) there are several ways to choose a basis
set of symmetric (permutationally invariant) operators, while only some
specific sets are convenient for studying macroscopic properties of large
number of qudits.

In the present paper we develop a general framework for the analysis of
global characteristics of multipartite qudit systems. We will: a) construct
projected $\tilde{Q}$-functions for $N$ qudit systems, containing full
macroscopic information; b) introduce a set of collective observables, which
characterizes macroscopic properties of qudits and compatible with the
phase-space description; c) propose two discrete collective tomographic
protocols (for general and symmetric $N$-qudit states) and provide an
explicit (partial) reconstruction expression for the $N$-qudit density
matrix in terms of average values of correlation functions of $\sim$ $%
N^{d^{2}-1}$ appropriate collective qudit operators. The example of $N$
qutrits will be analyzed in details and compared to $N$-qubit systems.

\section{N qudit projected $\widetilde{Q}$-function}

Let us consider $N$-qudit Hilbert space $\mathcal{H}_{d^{N}}=\mathcal{H}%
_{d}^{\otimes N}$, where $d$ is a prime number, spanned by the computational
basis $\left\vert \lambda\right\rangle =|l_{1},...,l_{N}\rangle$, $l_{i}\in%
\mathbb{Z}_{d}$. The standard unitary $N$-qudit operators are defined
according to \cite{simplect},\cite{stabilizers}

\begin{equation}
Z_{\alpha }=\sum_{\lambda }\omega ^{\lambda \alpha }\left\vert \lambda
\right\rangle \left\langle \lambda \right\vert ,\qquad X_{\beta
}=\sum_{\lambda }\left\vert \lambda +\beta \right\rangle \left\langle
\lambda \right\vert ,  \label{Nqudits}
\end{equation}%
where
\begin{equation}
\alpha =(a_{1},...,a_{N}),\qquad \beta =(b_{1},...,b_{N}),\qquad
a_{i},b_{i}=0,...,d-1,  \label{d string}
\end{equation}%
are $d$ - strings with elements from $\mathbb{Z}_{d}$, $\alpha \beta
=a_{1}b_{1}+...+a_{N}b_{N}$, $\omega =\exp \left( 2\pi i/d\right) $ sums and
multiplications are taken by mod $d$. The operators (\ref{Nqudits}) are
factorized into tensor products,

\begin{equation}
Z_{\alpha}={\displaystyle \bigotimes_{i=1}^{N}Z_{i}^{a_{i}},\qquad X_{\beta}=%
{\displaystyle \bigotimes_{i=1}^{N}X_{i}^{b_{i}},\qquad}}  \label{ZX}
\end{equation}

\noindent of single qudit Pauli operators \cite{Schwinger}

\begin{equation}
Z_{i}=\sum_{l=0}^{d-1}\omega^{l}\left\vert l\right\rangle \left\langle
l\right\vert ,\qquad X_{i}=\sum_{l=0}^{d-1}\left\vert l+1\right\rangle
\left\langle l\right\vert ,  \label{1qudit}
\end{equation}
and satisfy the commutation relation

\begin{equation}
Z_{\alpha}X_{\beta}=\omega^{\alpha\beta}X_{\beta}Z_{\alpha}.  \label{conmuta}
\end{equation}

Operators acting in $N$-qudit Hilbert space can be mapped into functions
labeled by a pair of $N$-tuples $\left( \alpha ,\beta \right) $. Two
mutually dual maps, known as discrete $Q$-symbols and $P$-symbols \cite{DCS}%
, \cite{WFpic}, \cite{Galetti}, have the form
\begin{eqnarray}
Q_{f}\left( \alpha ,\beta \right) &=&Tr\left( \hat{\Delta}^{(-1)}\left(
\alpha ,\beta \right) \hat{f}\right) ,\quad P_{f}\left( \alpha ,\beta
\right) =Tr\left( \hat{\Delta}^{(1)}\left( \alpha ,\beta \right) \hat{f}%
\right) ,  \label{symbols1} \\
\mathrm{Tr}\left( \hat{\Delta}^{(1)}(\alpha ,\beta )\hat{\Delta}%
^{(-1)}(\alpha ^{\prime },\beta ^{\prime })\right) &=&\delta _{\alpha \alpha
^{\prime }}\delta _{\beta \beta ^{\prime }},  \label{trac1}
\end{eqnarray}%
see (\ref{pmkernel}) for the explicit form of the kernels $\hat{\Delta}%
^{(\pm 1)}$. \ The first rank projectors
\begin{eqnarray}
\hat{\Delta}^{(-1)}\left( \alpha ,\beta \right) &=&|\alpha ,\beta \rangle
\langle \alpha ,\beta |,  \label{ab} \\
|\alpha ,\beta \rangle &=&Z_{\alpha }X_{\beta }\left\vert \xi \right\rangle
=\bigotimes_{i=1}^{N}|a_{i},b_{i}\rangle ,  \label{ab3}
\end{eqnarray}%
form an informationally complete set,
\begin{equation}
\sum_{\alpha ,\beta }\hat{\Delta}^{(-1)}\left( \alpha ,\beta \right) =d^{N}%
\hat{I}.  \label{Id}
\end{equation}%
The fiducial state $\left\vert \xi \right\rangle $ in (\ref{ab3}) is chosen
in a product form,

\begin{equation}
\left\vert \xi\right\rangle =\bigotimes_{i=1}^{N}\left\vert \xi\right\rangle
_{i},  \label{efd}
\end{equation}
in such a way that the single particle projectors on the states
\begin{equation}
|a_{i},b_{i}\rangle=Z_{i}^{a_{i}}X_{i}^{b_{i}}|\xi\rangle_{i},  \label{ab2}
\end{equation}
satisfy the symmetric informationally complete (SIC) POVM condition \cite%
{SIC}
\begin{equation}
\left\vert \left\langle
a_{i},b_{i}|a_{i}^{\prime},b_{i}^{\prime}\right\rangle \right\vert ^{2}=%
\frac{1+d\delta_{a_{i},a_{i}^{\prime}}\delta_{b_{i},b_{i}^{\prime}}}{1+d}.
\label{ecsp}
\end{equation}
An operator $\hat{f}\ $\ can be decomposed on the basis of $\hat{\Delta}%
^{(\pm1)}\left(\alpha,\beta\right)$ as
\begin{eqnarray}
\hat{f} & = & \sum_{\alpha,\beta}Q_{f}\left(\alpha,\beta\right)\hat{\Delta}%
^{(1)}\left(\alpha,\beta\right)  \label{fP} \\
& = & \sum_{\alpha,\beta}P_{f}\left(\alpha,\beta\right)\hat{\Delta}%
^{(-1)}\left(\alpha,\beta\right),
\end{eqnarray}
so that the average values are computed according to
\begin{equation}
\left\langle \hat{f}\right\rangle
=\sum_{\alpha,\beta}Q_{\rho}\left(\alpha,\beta\right)P_{f}\left(\alpha,\beta%
\right)=\sum_{\alpha,\beta}Q_{f}\left(\alpha,\beta\right)P_{\rho}\left(%
\alpha,\beta\right).  \label{prpq}
\end{equation}
It follows from (\ref{ab}) and (\ref{efd}) that
\begin{equation}
\mathcal{P}\hat{\Delta}^{(\pm1)}\left(\alpha,\beta\right)\mathcal{%
P^{\dagger}=}\hat{\Delta}^{(\pm1)}\left(\mathcal{\pi}\alpha,\mathcal{\pi}%
\beta\right)
\end{equation}
where $\mathcal{P}$ is the permutation operator and $\mathcal{\pi}\alpha$ is
a permutation of the string $\alpha$ corresponding to $\mathcal{P}$.

Then, $P$ and $Q$ - symbols of \textit{permutationally invariant} operators,
$\hat{s}=\mathcal{P}\hat{s}\mathcal{P}^{\dagger }$, are symmetric functions
of their arguments, and thus depend only on the corresponding weights $%
\mathbf{h}\left( \alpha ,\beta \right) $ \cite{FF} (see also Appendix A)
\begin{equation}
P_{s}\left( \alpha ,\beta \right) =\mathrm{Tr}\left( \hat{\Delta}%
^{(1)}(\alpha ,\beta )\hat{s}\right) \equiv P_{s}\left( \mathbf{h}\left(
\alpha ,\beta \right) \right) .  \label{P}
\end{equation}%
The weights $h\left( \alpha \right) $, being invariant under permutations
characteristics of $d$-strings (\ref{d string}), $h\left( \alpha \right)
=h\left( \mathcal{\pi }\alpha \right) $, are defined according to
\begin{equation}
h\left( \alpha \right) =\sum_{i=1}^{N}a_{i}=\sum_{k\in \mathbb{Z}%
_{d}}k\sum_{i=1}^{N}\delta _{a_{i},k},\;0\leq h(\alpha )\leq (d-1)N,
\end{equation}%
and can be arranged in a vector

\begin{equation}
\mathbf{\mathbf{h}}\left(\alpha,\beta\right)\mathbf{=}\{h(k\alpha+l\beta);%
\;k,l\in\mathbb{Z}_{d}\}.  \label{hh}
\end{equation}

\noindent The $d^{2}-1$\ components of the vector $\mathbf{h}%
\left(\alpha,\beta\right)$ form a basis in the space of symmetric functions
constructed on $(\alpha,\beta)$.

For instance, in case of qubits the three-dimesional $\mathbf{h}$ - vector
has the form

\begin{equation}
\mathbf{h}\left( \alpha ,\beta \right) =\{h(\alpha ),h(\beta ),h(\alpha
+\beta )\},
\end{equation}%
while for qutrits $\mathbf{h}$ - vector is eight-dimensional,%
\begin{equation}
\mathbf{h}\left( \alpha ,\beta \right) =\{h(\alpha ),h\left( 2\alpha \right)
,h(\beta ),h(2\beta ),h(\alpha +\beta ),h(2\alpha +2\beta ),h(2\alpha +\beta
),h(\alpha +2\beta )\}.
\end{equation}%
The points of two-dimensional grid $(\alpha \mathbf{,}\beta )$ can be
partially ordered according to the values of the weights (\ref{hh}) and an
arbitrary function constructed on $\left( \alpha ,\beta \right) $ satisfy
the following summation rule
\begin{equation}
\sum_{\alpha ,\beta }f\left( \alpha ,\beta \right) =\sum\limits_{\mathbf{m}%
}\sum_{\alpha ,\beta }\delta _{\mathbf{h}\left( \alpha ,\beta \right) ,%
\mathbf{m}}f\left( \alpha ,\beta \right) ,  \label{SM}
\end{equation}%
where the components of the vector $\mathbf{m}$ take values from $0$ to $%
(d-1)N$.

Taking into account (\ref{P}) and the summation rule (\ref{SM}) we observe
that the average value of any permutationally invariant operator in an
\textit{arbitrary} (not necessarily symmetric) state is computed according
to
\begin{equation}
\left\langle \hat{s}\right\rangle
=\sum_{\alpha,\beta}P_{s}(\alpha,\beta)Q_{\rho}(\alpha,\beta)=\sum_{\mathbf{m%
}}P_{s}\left(\mathbf{m}\right)\widetilde{Q}_{\rho}\left(\mathbf{m}\right),
\label{sAv}
\end{equation}
where
\begin{eqnarray}
\widetilde{Q}_{\rho}\left(\mathbf{m}\right) & = &
\sum_{\alpha,\beta}Q_{\rho}\left(\alpha,\beta\right)\delta_{\mathbf{h}%
\left(\alpha,\beta\right),\mathbf{m}}=Tr(\hat{\Delta}^{(-1)}(\mathbf{m})\hat{%
\rho}).  \label{Qp} \\
\hat{\Delta}^{(\pm1)}(\mathbf{m}) & = & \sum_{\alpha,\beta}\hat{\Delta}%
^{(\pm1)}\left(\alpha,\beta\right)\delta_{\mathbf{h}\left(\alpha,\beta%
\right),\mathbf{m}}.  \label{Qp2}
\end{eqnarray}
For symmetric states, $\mathcal{P}\hat{\rho}_{s}\mathcal{P}^{\dagger}=\hat{%
\rho}_{s}$, where $Q_{\rho_{s}}(\alpha,\beta)\equiv Q_{\rho_{s}}(\mathbf{h}%
\left(\alpha,\beta\right))$, one has
\begin{eqnarray}
\widetilde{Q}_{\rho_{s}}\left(\mathbf{m}\right) & = & Q_{\rho_{s}}(\mathbf{h}%
\left(\alpha,\beta\right))R_{\mathbf{m}}^{(d)},  \label{Qs} \\
R_{\mathbf{m}}^{(d)} & = & \sum_{\alpha,\beta}\delta_{\mathbf{h}%
\left(\alpha,\beta\right),\mathbf{m}}=\sum_{\alpha,\beta}\prod_{\substack{ %
k,l=0  \\ \{k,l\}\neq\{0,0\}  \\  \\ {d-1} \\ h(k\alpha}}^{d-1}\delta_{h(k%
\alpha+l\beta),m_{kl}},  \label{Rm}
\end{eqnarray}
where $R_{\mathbf{m}}^{(d)}$ is the multiplicity of each particular set
\begin{equation}
\mathbf{m}=\{m_{kl}=h(k\alpha+l\beta),\;0\leq
m_{kl}\leq(d-1)N,\;k,l=0,...,d-1\}.  \label{mkl}
\end{equation}
i.e. is the number of pairs $\left(\alpha,\beta\right)$ of $d$-strings
characterized by the same vector $\mathbf{m}=\{m_{kl}=h(k\alpha+l\beta),%
\;k,l=0,...,d-1\}$.

It follows from (\ref{sAv}) that the $\widetilde{Q}_{\rho }\left( \mathbf{m}%
\right) $-function contains complete and non-redundant information about all
macroscopic properties of the state $\hat{\rho}$ and can be considered as a
discrete distribution in $d^{2}-1$ dimensional \textit{macroscopic }%
measurement space $\mathcal{M}$ spanned by the vectors (\ref{mkl}). The
total number of multiplets $\{\mathbf{m=h}\left( \alpha ,\beta \right) \}$
(points in $\mathcal{M}$) is
\begin{equation}
\mathcal{N_{M}}=\frac{\left( N+d^{2}-1\right) !}{\left( d^{2}-1\right) !N!},
\label{Nm}
\end{equation}%
which is the amount of collective measurements fully determining the $%
\widetilde{Q}_{\rho }\left( \mathbf{m}\right) $ - function. In other words,
the global variables \textquotedblleft see\textquotedblright\ an $N$-qudit
state in form of $\widetilde{Q}_{\rho }\left( \mathbf{m}\right) $
distribution in $\mathcal{M}$. In the $N$-qubit case, the $\widetilde{Q}$%
-function has been extensively studied in \cite{macros}, \cite{galo} and
applied to the analysis of pure state thermalization effect and quantum
phase transitions \cite{Qapp}.

In the macroscopic limit $N\gg d$,
\begin{equation}
\mathcal{N_{M}}\sim N^{d^{2}-1}/\left( d^{2}-1\right) !,  \label{NmL}
\end{equation}%
which is significantly smaller than the number of points in the full
discrete phase-space $\sim d^{2N}$. In practice, distributions corresponding
to physically relevant macroscopic states tend to smooth shapes located in
certain areas of the measurement space $\mathcal{M}$, as it was observed in
qubit case \cite{macros}, \cite{galo}, and shown in the following examples.

\textit{Examples.}

1. For the fiducial state (\ref{efd}) - (\ref{ecsp}) one can easily obtain
\begin{equation}
\widetilde{Q}_{\xi }^{(d)}\left( \mathbf{m}\right) =\left( d+1\right) ^{-%
\frac{2}{d^{2}\left( d-1\right) }\sum_{k,l}m_{kl}}R_{\mathbf{m}}^{(d)},
\end{equation}%
which is a localized distribution (of size $\sim \sqrt{N}$) in the $\mathcal{%
M}$ - space, which tends to the Gaussian shape for $N\gg d$, for instance
(see Appendix C). For instance, in N-qubit case $\widetilde{Q}_{\xi }\left(
\mathbf{m}\right) $ is a\ single "spherically symmetric" Gaussian function
centered at $\mathbf{m}_{0}\mathbf{=(}3N/8,3N/8,3N/8)$,%
\begin{equation}
\widetilde{Q}_{\xi }^{(2)}\left( \mathbf{m}\right) \sim \exp (-2\frac{\left[
(m_{01}-\frac{3N}{8})^{2}+(m_{01}-\frac{3N}{8})^{2}+(m_{11}-\frac{3N}{8})^{2}%
\right] }{N},
\end{equation}%
while in N-qutri case $\widetilde{Q}_{\xi }\left( \mathbf{m}\right) $ takes
a form of a "squeezed" Gaussian, which can be compactly represented as
follows
\begin{gather}
, \\
\widetilde{Q}_{\xi }^{(3)}\left( \mathbf{m}\right) \sim \exp
(-\sum_{i=0}^{1}\sum_{j=0}^{i+1}\frac{\left[ (m_{ij}-\frac{8N}{9}%
)^{2}+(m_{2i2j}-\frac{8N}{9})^{2}-(m_{ij}-\frac{8N}{9})(m_{2i2j}-\frac{8N}{9}%
)\right] }{N}).
\end{gather}

2. For the $GHZ$-like $N$-qudit state

\begin{equation}
\left\vert GHZ\right\rangle =\frac{1}{\sqrt{d}}\sum_{l=0}^{d-1}\left\vert
l...l\right\rangle .
\end{equation}
one obtains by using (\ref{ab3}) for the discrete $Q$-function (\ref%
{symbols1}) the following explicitly invariant under particle permutations
expression

\begin{equation}
Q\left( \alpha ,\beta \right) =\left\vert \langle GHZ|\alpha ,\beta \rangle
\right\vert ^{2}=\frac{1}{d}\left\vert \sum_{l=0}^{d-1}\omega
^{l\sum_{i}a_{i}}\prod_{i=1}^{N}c_{l-b_{i}}^{\left( i\right) }\right\vert
^{2},  \label{GHZ}
\end{equation}%
where $a_{i}$, $b_{i}$ are components of the $d-$string (\ref{d string}) and
$c_{l}^{\left( i\right) }$ are the expansion coefficients of the $i$-th
particle fiducial state in the computational basis,
\begin{equation}
\left\vert \xi \right\rangle _{i}=\sum_{l=0}^{d-1}c_{l}^{\left( i\right)
}\left\vert l\right\rangle _{i},  \label{ci}
\end{equation}%
so that
\begin{equation}
\;\left\vert \xi \right\rangle =\sum_{\lambda }c_{\lambda }\left\vert
\lambda \right\rangle ,\;\left\vert \lambda \right\rangle =\left\vert
l_{1},..,l_{N}\right\rangle ,\;c_{\lambda }=\prod_{i=1}^{N}c_{l_{i}}^{\left(
i\right) },  \label{ciN}
\end{equation}%
Thus, according to (\ref{Qs}), we arrive at

\begin{equation}
\widetilde{Q}_{GHZ}^{(d)}\left(\mathbf{m}\right)=\frac{1}{d}\left\vert
\sum_{l=0}^{d-1}\omega^{m_{l0}}\prod_{i=1}^{N}c_{l-b_{i}}^{\left(i\right)}%
\right\vert ^{2}R_{\mathbf{m}}^{(d)},  \label{QGHZ}
\end{equation}
where the product $\prod_{i=1}^{N}c_{l-b_{i}}^{\left(i\right)}$ is a
function of $m_{0l}$, $l=1,2,..,d-1$ only.

For qubits, $d=2$, and the SIC POVM generating fiducial state
\begin{equation}
\left\vert \xi \right\rangle _{i}=\frac{|0\rangle _{i}+\zeta |1\rangle _{i}}{%
\sqrt{1+|\zeta |^{2}}},\quad \zeta =\frac{\sqrt{3}-1}{\sqrt{2}}e^{i\pi /4},
\label{ecd2}
\end{equation}%
the expression (\ref{QGHZ}) is reduced to

\begin{equation}
\widetilde{Q}_{GHZ}^{(2)}\left( \mathbf{m}\right) =\frac{1}{2\left(
1+\left\vert \zeta \right\vert ^{2}\right) ^{N}}\left\vert \zeta
^{m_{01}}+\left( -1\right) ^{m_{10}}\zeta ^{N-m_{01}}\right\vert ^{2}R_{%
\mathbf{m}}^{(2)},
\end{equation}%
where $R_{\mathbf{m}}^{(2)}$ is given in (\ref{R2}). The distribution $%
\widetilde{Q}_{GHZ}^{(2)}\left( \mathbf{m}\right) $ has a form of two
separated by a distance $\sim N$ along the axis $m_{01}$ discrete clouds,
each of size $\sim \sqrt{N}$ \cite{macros}, \cite{galo} in three-dimensional
measurement space. Each cloud acquires a Gaussian shape centered at $%
m_{01}=N(1\pm 1/\sqrt{3})/2$ in the limit $N\gg 1$. In Fig.1 we plot the
projection of $\widetilde{Q}_{GHZ}^{(2)}\left( \mathbf{m}\right) $ on the
axis $m_{01}$. there are two maximums at $m_{01}=N(1\pm 1/\sqrt{3})/2$

\begin{figure}[tbp]
\centering \includegraphics[scale=0.3]{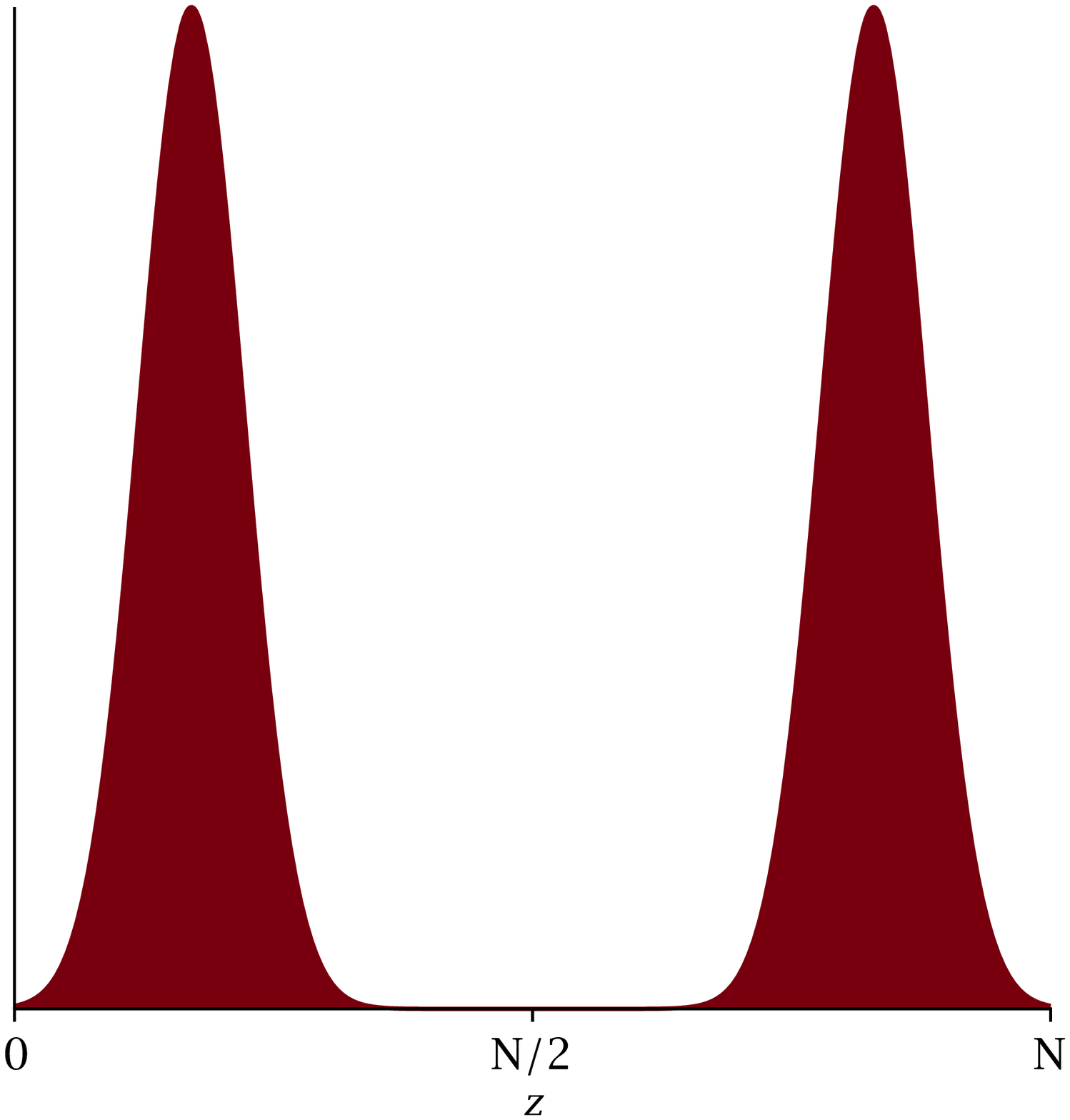}
\caption{Projection of $\widetilde{Q}_{GHZ}^{(2)}\left( \mathbf{m}\right) $
on the axis $m_{01}$. }
\end{figure}

For qutrits, $d=3$, one obtains

\begin{eqnarray}
\widetilde{Q}_{GHZ}^{(3)}\left( \mathbf{m}\right) &=&\frac{1}{2^{N}3}%
\left\vert 2^{m_{02}/3-m_{01}/6}\delta
_{m_{0,1},2m_{02}}+2^{m_{01}/3-m_{02}/6}\left( -\omega \right)
^{m_{01}-N}\delta _{2m_{01},m_{02}}\right.  \notag \\
&&\left. +2^{N/2-m_{01}/6-m_{02}/6}\left( -\omega \right)
^{m_{01}-m_{20}}\delta _{3N,m_{01}+m_{02}}\right\vert ^{2}R_{\mathbf{m}%
}^{(3)},
\end{eqnarray}%
where the fiducial state is chosen as
\begin{equation}
\left\vert \xi \right\rangle _{i}=\frac{1}{\sqrt{2}}\left( |0\rangle
_{i}+e^{i\pi /3}|1\rangle _{i}\right) ,  \label{ecd3}
\end{equation}%
and $R_{\mathbf{m}}^{(3)}$ is given in (\ref{R3}). As it can be appreciated
from the above expression, $\widetilde{Q}_{GHZ}^{(3)}\left( \mathbf{m}%
\right) $ is a superposition of three localized clusters, each tending to a
Gaussian form centered at in the plane $\left( m_{01},m_{02}\right) $ of
eight-dimensional $\mathcal{M}$ - space at $\left( m_{01},m_{02}\right)
=\left( N,N/2\right) ,\left( N/2,N\right) ,\left( 3N/2,3N/2\right) $.

\begin{figure}[tbp]
\centering \includegraphics[scale=0.5]{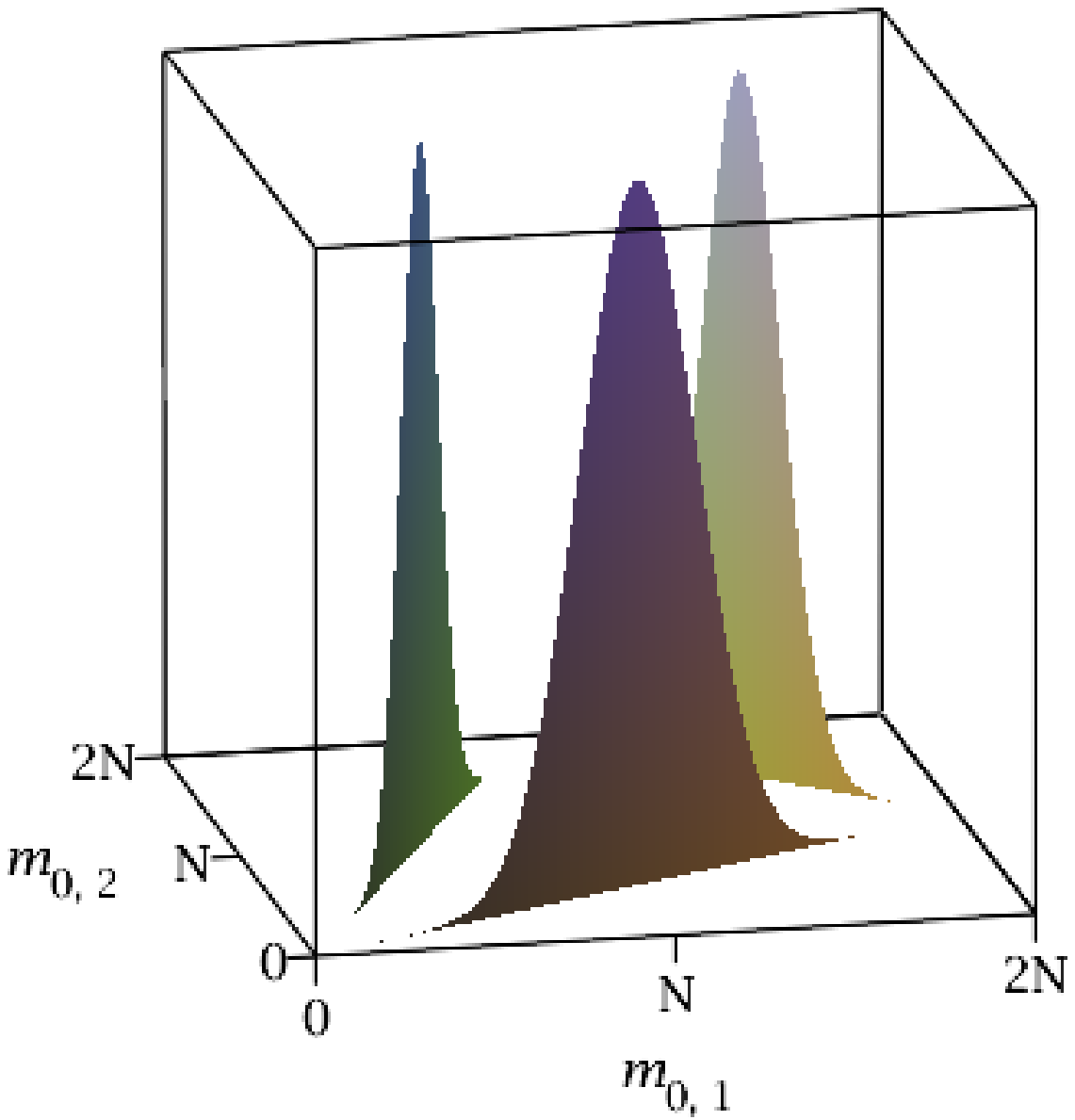}
\caption{Projection of $\widetilde{Q}_{GHZ}^{(3)}\left( \mathbf{m}\right) $
on the plane $m_{01}-m_{02}$.}
\end{figure}

\bigskip{}

In qubit case $\widetilde{Q}_{\rho }\left( \mathbf{m}\right) $ can be
plotted as density distributions in a three-dimensional space. For higher
dimensions $\widetilde{Q}_{\rho }\left( \mathbf{m}\right) $ can be
represented only in form of projections into hyper-planes in the full $%
\mathcal{M}$ - space. Nevertheless, the analytical properties of the $%
\widetilde{Q}$-functions are very useful for analysis of the global
properties of macroscopic systems, as it will be shown below.

\subsection{Collective operators}

\noindent Let us introduce the following set of collective (invariant under
particle permutations) Hermitian operators for an $N$-qudit system
\begin{equation}
\hat{O}_{k,l}=N\hat{I}-\frac{2}{\left(d-1\right)d^{N}}\sum_{\alpha,\beta}h%
\left(k\alpha+l\beta\right)\left\vert \alpha,\beta\right\rangle \left\langle
\alpha,\beta\right\vert ,\quad k,l=0,...,d-1.  \label{O1}
\end{equation}
Their $P$-symbols (\ref{symbols1}) are proportional to the corresponding
weights:

\begin{equation}
P_{\hat{O}_{kl}}=\frac{1}{d^{N}}\left[N-\frac{2}{d-1}h\left(k\alpha+l\beta%
\right)\right],  \label{PO}
\end{equation}
which allows to associate the axes in the measurement space $\mathcal{M}$
with the collective operators (\ref{O1}). This can be very helpful for
determination of measurements required for a detection of quantum states.
For instance, the main features of the qutrit GHZ-like state (\ref{GHZ}) can
be recognized by measuring $\hat{O}_{0,1\text{ }}$and $\hat{O}_{0,2\text{ }}$%
operators and their moments, see Fig.2.

It is proven in Appendix B that the operators (\ref{O1}) are split into $d+1$
disjoint sets of $d$ commuting operators:
\begin{equation}
\{\hat{O}_{\lambda k,\lambda l},~\lambda =1,...,d-1\},~~\{\hat{O}%
_{0,l}\},~~\{\hat{O}_{k,0}\},~k,l=1,...,d-1,
\end{equation}%
where
\begin{equation}
\lbrack \hat{O}_{\lambda k,\lambda l},\hat{O}_{k,l}]=0,~~[\hat{O}_{0,l},\hat{%
O}_{0,l^{\prime }}]=0,~~[\hat{O}_{k,0},\hat{O}_{k^{\prime },0}]=0,
\label{Op1}
\end{equation}%
and
\begin{equation}
tr\left( \hat{O}_{k,l}\hat{O}_{k^{\prime },l^{\prime }}\right) =0,\quad
k^{\prime }l\neq kl^{\prime }.  \label{Op2}
\end{equation}%
The collective observables (\ref{O1}) can be represented in form of a direct
product of single-particle operators
\begin{eqnarray}
\hat{O}_{k,l} &=&\sum_{i=1}^{N}\hat{I}\otimes ...\otimes \hat{O}%
_{k,l}^{(i)}\otimes ...\otimes \hat{I},  \label{Okl} \\
\hat{O}_{k,l}^{(i)} &=&\hat{I}^{(i)}-\frac{2}{d\left( d-1\right) }%
\sum_{a_{i},b_{i}=0}^{d-1}\left\{ ka_{i}+lb_{i}\right\} \left\vert
a_{i},b_{i}\right\rangle \left\langle a_{i},b_{i}\right\vert ,  \label{Okli}
\end{eqnarray}%
where $\left\vert a_{i},b_{i}\right\rangle $ are the states (\ref{ab2}) and
the operations in $\{\cdot \}$ are taken mod $d$.\ The operators (\ref{Okli}%
) are normalized according to
\begin{equation}
Tr\hat{O}_{k,l}^{(i)}=0,\quad Tr\left( \left[ \hat{O}_{k,l}^{(i)}\right]
^{2}\right) =\frac{d}{3\left( d-1\right) },
\end{equation}%
and form a basis of the $su(d)$ algebra similar to that introduced in \cite%
{Patera}. The explicit form of the matrix elements of the operators $\hat{O}%
_{k,l}^{(i)}$ in the computational basis is given in Appendix B, Eqs. (\ref%
{OpqD})-(\ref{OpqND}).

Taking into account (\ref{ab2}) one obtains that $d-1$ operators $\hat{O}%
_{0,l}^{(i)}$ are diagonal in the logical basis (see Appendix B). The
elements of the commuting sets containing non-diagonal operators $\hat{O}%
_{k,l}^{(i)},k\neq0$, are convenient to label as $\hat{O}_{\lambda,\lambda
m}^{(i)}$, which matrix elements in the computational basis are given in
Appendix B. The commuting sets $\{\hat{O}_{\lambda,\lambda m}$, $[\hat{O}%
_{\lambda,\lambda m},\hat{O}_{\lambda,\lambda m^{\prime}}]=0$, $%
m,m^{\prime}=0,...,d-1\}$ are obtained from the diagonal set $\{\hat{O}%
_{0,l}\}$ by $SU(d)$ transformations, and thus can be efficiently measured.

\textit{Examples}

1. Qubits, $d=2.$ The fiducial state (\ref{ecd2}) leads to the natural
representation
\begin{equation}
\hat{O}_{0,1}^{(i)}=\frac{1}{\sqrt{3}}\sigma_{z}^{(i)},~~\hat{O}_{1,0}^{(i)}=%
\frac{1}{\sqrt{3}}\sigma_{x}^{(i)},~~\hat{O}_{1,1}^{(i)}=\frac{1}{\sqrt{3}}%
\sigma_{y}^{(i)},
\end{equation}
so that the operators (\ref{O1}) coincide up to a constant factor with the
spin collective operators (\ref{S}).

2. Qutrits, $d=3$. Taking the fiducial state (\ref{ecd3}) we obtain the
following four commutative sets of cyclic (up to a constant) operators, $%
\hat{O}_{k,l}^{(i)}=4\left(\hat{O}_{k,l}^{(i)}\right)^{3}$, (compare to \cite%
{Patera})
\begin{gather}
\quad\hat{O}_{0,1}^{(i)}=\frac{1}{2}\left[%
\begin{array}{ccc}
0 & 0 & 0 \\
0 & 1 & 0 \\
0 & 0 & -1%
\end{array}%
\right],\;\hat{O}_{0,2}^{(i)}=\frac{1}{2}\left[%
\begin{array}{ccc}
1 & 0 & 0 \\
0 & 0 & 0 \\
0 & 0 & -1%
\end{array}%
\right],  \label{o1} \\
\hat{O}_{1,0}^{(i)}=\frac{i}{2\sqrt{3}}\left[%
\begin{array}{ccc}
0 & -1 & 1 \\
1 & 0 & -1 \\
-1 & 1 & 0%
\end{array}%
\right],\;\hat{O}_{2,0}^{(i)}=\frac{1}{2\sqrt{3}}\left[%
\begin{array}{ccc}
0 & e^{-i\pi/6} & e^{i\pi/6} \\
e^{i\pi/6} & 0 & e^{-i\pi/6} \\
e^{-i\pi/6} & e^{i\pi/6} & 0%
\end{array}%
\right],  \label{o2} \\
\hat{O}_{1,1}^{(i)}=\frac{1}{2\sqrt{3}}\left[%
\begin{array}{ccc}
0 & -i & e^{-i5\pi/6} \\
i & 0 & e^{i\pi/6} \\
e^{i5\pi/6} & e^{-i\pi/6} & 0%
\end{array}%
\right],\;\hat{O}_{2,2}^{(i)}=\frac{1}{2\sqrt{3}}\left[%
\begin{array}{ccc}
0 & e^{-i\pi/6} & e^{i5\pi/6} \\
e^{i\pi/6} & 0 & i \\
e^{-i5\pi/6} & -i & 0%
\end{array}%
\right],  \label{o3} \\
\hat{O}_{2,1}^{(i)}=\frac{1}{2\sqrt{3}}\left[%
\begin{array}{ccc}
0 & e^{-i\pi/6} & -i \\
e^{i\pi/6} & 0 & e^{-i5\pi/6} \\
i & e^{i5\pi/6} & 0%
\end{array}%
\right],\;\hat{O}_{1,2}^{(i)}=\frac{1}{2\sqrt{3}}\left[%
\begin{array}{ccc}
0 & -i & e^{-i\pi/6} \\
i & 0 & e^{i5\pi/6} \\
e^{i\pi/6} & e^{-i5\pi/6} & 0%
\end{array}%
\right].  \label{o4}
\end{gather}
The sets (\ref{o2}) - (\ref{o4}) can be obtained form the set (\ref{o1}) by $%
SU(3)$ rotations.

\section{Collective tomography}

By a direct substitution one can prove that the symmetrized operators (\ref%
{Qp2}) form a bi-orthogonal set
\begin{equation}
Tr\left[\hat{\Delta}^{\left(1\right)}\left(\mathbf{m}\right)\hat{\Delta}%
^{\left(-1\right)}\left(\mathbf{m}^{\prime}\right)\right]=R_{\mathbf{m}%
}^{(d)}\delta_{\mathbf{m},\mathbf{m}^{\prime}}.  \label{biort}
\end{equation}
This suggests to approximate the density matrix by ``inverting''\ Eq. (\ref%
{Qp}) in the form similar to Eq.(\ref{fP})
\begin{equation}
\hat{\rho}\approx\hat{\rho}_{rec}=\sum\limits _{\mathbf{m}}\widetilde{Q}%
_{\rho}\left(\mathbf{m}\right)\left(R_{\mathbf{m}}^{(d)}\right)^{-1}\hat{%
\Delta}^{\left(1\right)}\left(\mathbf{m}\right).  \label{rho_rec}
\end{equation}
The above equation is a formal expression for the state reconstruction by
using the results of all possible collective measurements stored in $%
\widetilde{Q}_{\rho}\left(\mathbf{m}\right)$. The ``tomographic
representation''\ (\ref{rho_rec}) is incomplete since the map (\ref{Qp}) is
not faithful. In other words, Eq.(\ref{rho_rec}) should be considered as a
form of arranging the information obtained from $\sim N^{d^{2}-1}$
collective measurements (corresponding to the total number of multiplets \{$%
\mathbf{m}$\}) in a $d^{N}\times d^{N}$ matrix.

Equation (\ref{rho_rec}) can be explicitly rewritten in terms of average
values of symmetrized monomials (\ref{Nqudits}) as follows
\begin{eqnarray}
\hat{\rho}_{rec} & = & d^{-N}\sum\limits _{\mathbf{m}}\left(R_{\mathbf{m}%
}^{(d)}\right)^{-1}\left\langle \hat{D}_{\mathbf{m}}\right\rangle \hat{D}_{%
\mathbf{m}}^{\dagger},  \label{rho1} \\
\hat{D}_{\mathbf{m}} & = & \sum_{\alpha,\beta}\delta_{\mathbf{h}%
\left(\alpha,\beta\right),\mathbf{m}}Z_{\alpha}X_{\beta},\quad\left\langle
\hat{D}_{\mathbf{m}}\right\rangle =Tr\left(\hat{\rho}\hat{D}_{\mathbf{m}%
}\right),  \label{D}
\end{eqnarray}
which is a generalization of results obtained in \cite{ariana}. The
operators (\ref{D}) can be always expressed in terms of polynomials of the
collective operators (\ref{O1})
\begin{equation}
\hat{D}_{\mathbf{m}}=\sum_{\mathbf{\mathtt{p}}}c_{\mathbf{\mathtt{p}}%
}^{\left(\mathbf{m}\right)}\prod_{k,l}O_{k,l}^{\mathtt{p}_{k,l}},\quad%
\sum_{k,l}\mathtt{p}_{k,l}\leq\sum_{k,l}m_{kl},\;
\end{equation}
(see Appendix D), and thus, accessed from Von Neumann measurements.

Since any state obtained from $\hat{\rho}$ by particle permutations, $%
\mathcal{P}\hat{\rho}\mathcal{P^{\dagger}}$, leads to the same $\hat{\rho}%
_{rec}$, the density matrix reconstructed according to Eq.(\ref{rho1}) is
related to the original one by the full symmetrization,
\begin{equation}
\hat{\rho}_{rec}=\frac{1}{N!}\sum_{\mathcal{P}}\mathcal{P}\hat{\rho}\mathcal{%
P^{\dagger}}.
\end{equation}
Thus, the reconstruction (\ref{rho1}) is exact for permutationally invariant
(symmetric) states. It is straightforward to obtain a closed from expression
for the reconstruction fidelity of a pure state $\hat{\rho}=\left\vert
\psi\right\rangle \left\langle \psi\right\vert $,

\begin{equation}
\mathcal{F}=Tr\left(\hat{\rho}\hat{\rho}_{rec}\right)=\frac{1}{N!}\sum_{%
\mathcal{P}}\left\vert \left\langle \psi\right\vert \mathcal{P}\left\vert
\psi\right\rangle \right\vert ^{2},
\end{equation}
which reaches its minimum value

\begin{equation}
\mathcal{F}_{min}=\frac{1}{N!},  \label{Fmin}
\end{equation}
for states that become orthogonal under any particle permutation, e.g for $N$%
-qudit, $d\geq N$, elements of the computational basis $\left\vert
\lambda\right\rangle =\left\vert l_{1},...,l_{N}\right\rangle $, with $%
l_{i}\neq l_{j}$.

For mixed states the minimum value of the fidelity can be substantiality
smaller then given in Eq.(\ref{Fmin}). For instance, for 2 and 3 qubits one
obtains $\mathcal{F}_{\min}\mathcal{=}1/2$ and $\mathcal{F}_{\min}\mathcal{=}%
1/8$ correspondingly; for 2 and 3 qutrits $\mathcal{F}_{\min}\mathcal{=}1/9$
and $\mathcal{F}_{\min}\mathcal{=}1/27$.

\subsection{Symmetric space reconstruction}

In the particular case of fully symmetric states, the reconstruction
expressions (\ref{rho_rec})-(\ref{rho1}) become exact and can be rewritten
in an explicit form by projecting the expansion (\ref{fP}) into the
symmetric subspace.

Let us consider a symmetric density matrix, i.e. $\hat{\rho}_{s}=\hat{\Pi}%
_{s}\hat{\rho}_{s}\hat{\Pi}_{s}$, where $\hat{\Pi}_{s}$ is the projection
operator on the symmetric subspace $\mathcal{H}_{sym}$ of $N$ qudits
\begin{equation}
\hat{\Pi}_{s}=\sum_{\mathbf{\mathfrak{p}}}\left\vert \mathbf{\mathfrak{p}}%
;N\right\rangle \left\langle \mathbf{\mathfrak{p}};N\right\vert ,
\end{equation}%
and permutationally invariant states
\begin{eqnarray}
\left\vert \mathbf{\mathfrak{p}};N\right\rangle &=&\left\vert
p_{1},...p_{d-1};N\right\rangle ,\;  \label{DS} \\
p_{j} &=&0,...,N,\;~~\sum_{j=1}^{d-1}p_{j}\leq N,
\end{eqnarray}%
are elements of an orthonormal basis, $\langle \mathbf{\mathfrak{p}}^{\prime
},N\left\vert \mathbf{\mathfrak{p}},N\right\rangle =\delta _{\mathbf{%
\mathfrak{p}}^{\prime },\mathbf{\mathfrak{p}}}$, in $d_{sym}$-dimensional
Hilbert space $\mathcal{H}_{sym}$,
\begin{equation}
d_{sym}=\frac{\left( N+d-1\right) !}{\left( d-1\right) !N!}.
\end{equation}%
The states (\ref{DS}) are expanded in the computational basis according to
\begin{equation}
\left\vert \mathbf{\mathfrak{p}},N\right\rangle =\mathcal{N}_{\mathbf{%
\mathfrak{p}}}\sum_{\lambda }\prod_{i=1}^{d-1}\delta _{\eta _{i}^{(\lambda
)},p_{i}}\left\vert \lambda \right\rangle ,
\end{equation}%
where
\begin{equation}
\eta _{i}^{(\lambda )}=\frac{\left( -1\right) ^{i+1}}{\left( d-1-i\right) !i!%
}\sum_{j=1}^{N}\prod_{\substack{ k=0  \\ k\neq i}}^{d-1}(k-l_{j}),
\label{nu}
\end{equation}%
denotes the number of coefficients $l_{j}$ equal to $i=1,...,d-1$ in the
state $\left\vert \lambda \right\rangle =\left\vert
l_{1},...,l_{N}\right\rangle $, and
\begin{equation}
\mathcal{N}_{\mathbf{\mathfrak{p}}}=\sqrt{\frac{%
p_{1}!...p_{d-1}!(N-p_{1}-...-p_{d-1})!}{N!}},
\end{equation}%
is the normalization constant. It is easy to see that $\eta _{i}^{(\lambda
)} $ are symmetric functions of $\lambda $, $\eta _{i}^{(\lambda )}=\eta
_{i}(h(\lambda ))$.

A density matrix $\hat{\rho}_{s}$ can be represented according to Eq. (\ref%
{fP}) as follows

\begin{equation}
\hat{\rho}_{s}=\sum_{\alpha ,\beta }tr\left( \hat{\rho}_{s}\hat{\Delta}%
_{s}^{(-1)}\left( \alpha ,\beta \right) \right) \hat{\Delta}_{s}^{(1)}\left(
\alpha ,\beta \right) ,  \label{rhos}
\end{equation}%
where $\hat{\Delta}_{s}^{(\pm 1)}\left( \alpha ,\beta \right) $ are the
kernels (\ref{pmkernel}) projected into the symmetric subspace,
\begin{equation}
\hat{\Delta}_{s}^{(\pm 1)}\left( \alpha ,\beta \right) =\hat{\Pi}_{s}\hat{%
\Delta}^{(\pm 1)}\left( \alpha ,\beta \right) \hat{\Pi}_{s}\equiv \hat{\Delta%
}_{s}^{(\pm 1)}\left( \mathbf{h}(\alpha ,\beta )\right) ,  \label{Dsym1}
\end{equation}%
acting in $\mathcal{H}_{sym}$.

The first-rank projectors
\begin{equation}
\hat{\Delta}_{s}^{(-1)}\left(\mathbf{h}(\alpha,\beta)\right)=\left\vert
\phi_{\mathbf{h}\left(\alpha,\beta\right)}\right\rangle \left\langle \phi_{%
\mathbf{h}\left(\alpha,\beta\right)}\right\vert ,~  \label{Qsym1}
\end{equation}
where

\begin{eqnarray}
\left\vert \phi _{\mathbf{h}\left( \alpha ,\beta \right) }\right\rangle &=&%
\hat{\Pi}_{s}\left\vert \alpha ,\beta \right\rangle =\sum_{\mathbf{\mathfrak{%
p}}}\mathcal{N}_{\mathbf{\mathfrak{p}}}\Upsilon _{\mathbf{\mathfrak{p}}}(%
\mathbf{h}\left( \alpha ,\beta \right) )\left\vert \mathbf{\mathfrak{p}}%
,N\right\rangle ,  \label{absym1} \\
\Upsilon _{\mathbf{\mathfrak{p}}}(\mathbf{h}\left( \alpha ,\beta \right) )
&=&\sum_{\lambda }\omega ^{\alpha \lambda }c_{\lambda -\beta
}\prod_{i=1}^{d-1}\delta _{\eta _{i}^{(\lambda )},p_{i}},
\end{eqnarray}%
being $c_{\mu }$ the fiducial state expansion coefficients (\ref{ciN}),
define the measurement sets and satisfy the completeness condition
\begin{equation}
\sum\limits_{\mathbf{m}}\hat{E}_{\mathbf{m}}^{(d)}=\hat{\Pi}_{s},\quad \hat{E%
}_{\mathbf{m}}^{(d)}=d^{-N}R_{\mathbf{m}}^{(d)}\hat{\Delta}_{s}^{(-1)}\left(
\mathbf{h}(\alpha ,\beta )=\mathbf{m}\right) ,  \label{complete}
\end{equation}%
where $R_{\mathbf{m}}^{(d)}$is defined in (\ref{Rm}) and $\hat{\Pi}_{s}$
acts as an identity operator in $\mathcal{H}_{sym}$. After a straightforward
algebra one can show that the $\hat{\Delta}_{s}^{(1)}$- kernel (\ref{Dsym1})
in the basis (\ref{DS}) acquires the form
\begin{equation}
\left\langle \mathfrak{\mathbf{\mathfrak{t}}},N\right\vert \hat{\Delta}%
_{s}^{(1)}\left( \mathbf{m}\right) \left\vert \mathbf{\mathfrak{t}}^{\prime
},N\right\rangle =\sum_{\mathbf{\mathfrak{t},\mathfrak{t^{\prime }}}}%
\mathcal{N}_{\mathbf{\mathfrak{t}}}\mathcal{N}_{\mathfrak{\mathbf{\mathfrak{%
t^{\prime }}}}}\left[ \sum_{\mathbf{q}}g^{(d)}\left( \mathbf{q},\mathbf{m}%
\right) f^{(d)}(\mathbf{q})C_{\mathfrak{t,t^{\prime }}}(\mathbf{q})\right] ,
\end{equation}%
where
\begin{equation}
g^{(d)}\left( \mathbf{q},\mathbf{m}\right) =\sum_{\alpha ,\beta }\omega
^{\alpha \delta -\beta \gamma }\delta _{\mathbf{h}\left( \alpha ,\beta
\right) ,\mathbf{m}}\delta _{\mathbf{h}\left( \gamma ,\delta \right) ,%
\mathbf{q}},  \label{gpr}
\end{equation}%
\begin{equation}
C_{\mathfrak{t,t^{\prime }}}(\mathbf{q})=\sum_{\gamma ,\delta ,\lambda
}\omega ^{\gamma \lambda }\prod_{i=1}^{d-1}\delta _{\eta _{i}^{(\lambda
)},t_{i}}\prod_{j=1}^{d-1}\delta _{\eta _{j}^{(\lambda -\delta
)},t_{j}^{\prime }}\delta _{\mathbf{h}\left( \gamma ,\delta \right) ,\mathbf{%
q}},
\end{equation}%
are the special discrete functions (special functions of discrete variables)
and the coefficient
\begin{equation}
f^{(d)}(\mathbf{q})=\sum_{\gamma ,\delta }\left( \left\langle \xi
\right\vert Z_{\gamma }X_{\delta }\left\vert \xi \right\rangle \right)
^{-1}\delta _{\mathbf{h}\left( \gamma ,\delta \right) ,\mathbf{q}}=R_{%
\mathbf{q}}^{(d)}\left( \left\langle \xi \right\vert Z_{\gamma }X_{\delta
}\left\vert \xi \right\rangle \right) ^{-1}|_{\mathbf{h}\left( \gamma
,\delta \right) =\mathbf{q}},  \label{fp}
\end{equation}%
can be always evaluated analytically, see (\ref{ZXme}).

Finally, the explicit reconstruction expression in the symmetric subspace $%
\mathcal{H}_{sym}$ acquires the form

\begin{equation}
\hat{\rho}_{s}=d^{N}\sum_{\mathbf{m}}\sigma _{\mathbf{m}}\left( R_{\mathbf{m}%
}^{(d)}\right) ^{-1}\hat{\Delta}_{s}^{(1)}\left( \mathbf{m}\right) ,
\label{symrec}
\end{equation}%
where
\begin{equation}
\sigma _{\mathbf{m}}=Tr\left( \hat{E}_{\mathbf{m}}^{(d)}\hat{\rho}%
_{s}\right) ,~~\sum_{\mathbf{m}}\sigma _{\mathbf{m}}=1,  \label{psym}
\end{equation}%
are measured probabilities. The total number of projections required in this
protocol is given by Eq.(\ref{Nm}), while the density matrix of a fully
symmetric state contains at most $d_{sym}^{2}-1$ independent parameters.
Such a redundancy occurs because the probabilities (\ref{psym}) are not
linearly independent and satisfy the following self-consistence conditions
\begin{equation}
\sigma _{\mathbf{q}}=d^{N}\sum_{\mathbf{m}}\sigma _{\mathbf{m}}\left( R_{%
\mathbf{m}}^{(d)}\right) ^{-1}\left\langle \phi _{\mathbf{q}}\right\vert
\hat{\Delta}_{s}^{(1)}\left( \mathbf{m}\right) \left\vert \phi _{\mathbf{q}%
}\right\rangle .  \label{redundance}
\end{equation}%
In order to estimate the accuracy of the reconstruction scheme (\ref{symrec}%
) we numerically studied the minimum square error (MSE) of the quadratic
Hilbert-Schmidt distance between a real state $\hat{\rho}_{s}$ and its
estimate $\hat{\rho}_{est}$ according to the Crámer-Rao lower bound \cite%
{Helstrom} for qubit and qutrit states, see Appendix E.

1. In N-qubit case, $d=2$, a single index (\ref{nu}) defines symmetric
states,
\begin{equation}
\eta _{1}^{(\lambda )}=\sum_{i=1}^{N}l_{i}=h\left( \lambda \right) .
\end{equation}

The elements of the basis in the symmetric subspaces are the well known
Dicke states \cite{Dicke}
\begin{eqnarray}
\left\vert p_{1};N\right\rangle & = & \mathcal{N}_{p_{1}}\sum_{\lambda}%
\delta_{h\left(\lambda\right),p_{1}}\left\vert \lambda\right\rangle , \\
\mathcal{N}_{p_{1}} & = & \sqrt{\frac{p_{1}!\left(N-p_{1}\right)!}{N!}}.
\end{eqnarray}
The symmetrized discrete coherent states (\ref{absym1}) are expanded in the
Dicke basis as follows

\begin{eqnarray}
\left\vert \phi_{\mathbf{h}\left(\alpha,\beta\right)}\right\rangle & = &
\sum_{p_{1}}\mathcal{N}_{p_{1}}\Upsilon_{p_{1}}\left(\mathbf{h}%
\left(\alpha,\beta\right)\right)\left\vert p_{1};N\right\rangle , \\
\Upsilon_{p_{1}}\left(\mathbf{h}\left(\alpha,\beta\right)\right) & = &
\left(1+\left\vert \zeta\right\vert
^{2}\right)^{-N/2}\sum_{\lambda}(-1)^{\alpha\lambda}\zeta^{h\left(\lambda+%
\beta\right)}\delta_{h\left(\lambda\right),p_{1}}.
\end{eqnarray}
where $\zeta$ is defined in Eq. (\ref{ecd2}). The functions (\ref{gpr}) and (%
\ref{fp}) have the form

\begin{gather*}
g^{(2)}\left(\mathbf{q},\mathbf{m}\right)=\sum_{\alpha,\beta}(-1)^{\alpha%
\delta+\beta\gamma}\delta_{\mathbf{h}\left(\alpha,\beta\right),\mathbf{m}%
}\delta_{\mathbf{h}\left(\gamma,\delta\right),\mathbf{q}}, \\
f^{(2)}(\mathbf{q})=3^{\frac{q_{10}+q_{01}+q_{11}}{4}}i^{\frac{%
q_{11}-q_{10}-q_{01}}{2}}R_{\mathbf{q}}^{(2)},
\end{gather*}
where $h\left(\gamma\right)=q_{10},h\left(\delta\right)=q_{01},h\left(%
\gamma+\delta\right)=q_{11}$ and
\begin{equation}
C_{\mathfrak{t,t^{\prime}}}(\mathbf{q})=\sum_{\gamma,\delta,\lambda}\left(-1%
\right)^{\gamma\lambda}\delta_{h\left(\lambda\right),t_{1}}\delta_{h\left(%
\lambda-\delta\right),t_{1}^{\prime}}\delta_{h\left(\gamma\right),q_{10}}%
\delta_{h\left(\delta\right),q_{01}}\delta_{h\left(\gamma+\delta%
\right),q_{11}}
\end{equation}
can be expressed in terms of $_{4}F_{3}$ - \ functions.

We have numerically found that the MSE $\sqrt{\left\langle \left\langle
\mathcal{E}_{min}^{2}\right\rangle \right\rangle }$, averaged over 200 pure
and mixed states, is inversely proportional to the square root of the number
of trials $M$,
\begin{equation}
\sqrt{\left\langle \left\langle \mathcal{E}_{min}^{2}\right\rangle
\right\rangle }\approx \frac{\lambda }{\sqrt{M}}.  \label{Error}
\end{equation}%
In Fig. 3 we plot the proportionality constant $\lambda $ for pure and mixed
states for $N=1,...,6$ qubits and compare with the values corresponding to
the SIC POVM tomographic protocol \cite{SIC}.

\begin{figure}[tbp]
\centering \includegraphics[scale=0.45]{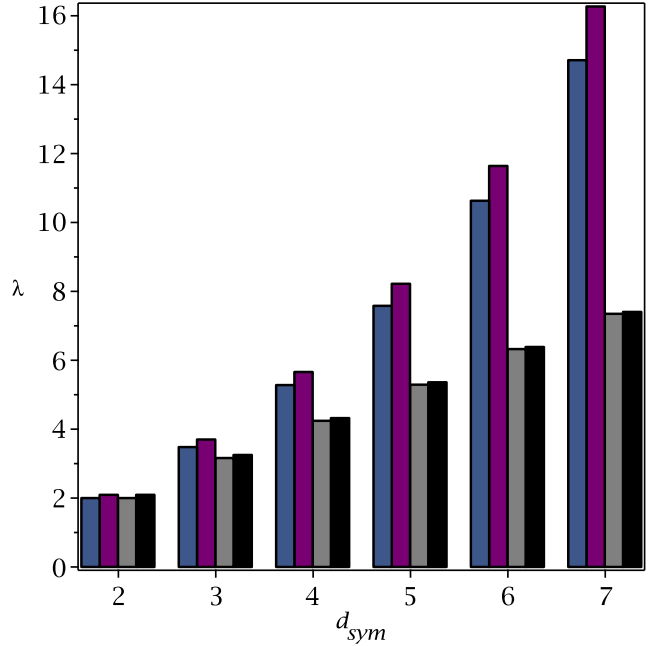}
\caption{First two columns: the proportionality constant $\protect\lambda$
appearing in the minimum square error (\protect\ref{Error}) for pure (left
column) and mixed (right column) states $d_{sym}=N+1$ for $N=1,...,6$ qubits
corresponding to the reconstruction skill (\protect\ref{symrec}); Second two
columns: the corresponding values of $\protect\lambda$ for SIC tomographic
protocol.}
\end{figure}

\bigskip {} 2. In N-qutrit case, $d=3$, there are two indexes (\ref{nu})
that define qutrit symmetric space,
\begin{eqnarray}
\eta _{1}^{(\lambda )} &=&\sum_{i=1}^{N}l_{i}\left( 2-l_{i}\right) =\frac{2}{%
3}\left[ h\left( 2\lambda \right) -\frac{1}{2}h\left( \lambda \right) \right]
=p_{1}, \\
\eta _{2}^{(\lambda )} &=&\sum_{i=1}^{N}\frac{l_{i}\left( l_{i}-1\right) }{2}%
=\frac{2}{3}\left[ h\left( \lambda \right) -\frac{1}{2}h\left( 2\lambda
\right) \right] =p_{2},
\end{eqnarray}%
thus, the basis in the symmetric subspace has the form

\begin{eqnarray}
\left\vert p_{1},p_{2};N\right\rangle &=&\mathcal{N}_{p_{1}p_{2}}\sum_{%
\lambda }\delta _{h\left( \lambda \right) ,p_{1}+2p_{2}}\delta _{h\left(
2\lambda \right) ,2p_{1}+p_{2}}\left\vert \lambda \right\rangle , \\
\mathcal{N}_{p_{1}p_{2}} &=&\sqrt{\frac{p_{1}!p_{2}!(N-p_{1}-p_{2})!}{N!}}.
\end{eqnarray}%
The symmetrized discrete coherent states (\ref{absym1}) are

\begin{eqnarray}
\left\vert \phi _{\mathbf{h}\left( \alpha ,\beta \right) }\right\rangle
&=&\sum_{p_{1},p_{2}}\mathcal{N}_{p_{1}p_{2}}\Upsilon _{p_{1}p_{2}}\left(
\mathbf{h}\left( \alpha ,\beta \right) \right) \left\vert
p_{1},p_{2};N\right\rangle , \\
\Upsilon _{p_{1}p_{2}}\left( \mathbf{h}\left( \alpha ,\beta \right) \right)
&=&\sum_{\lambda }\omega ^{\alpha \lambda }c_{\lambda -\beta }\delta
_{h\left( \lambda \right) ,p_{1}+2p_{2}}\delta _{h\left( 2\lambda \right)
,2p_{1}+p_{2}},
\end{eqnarray}%
where $\omega =\exp (2\pi i/3)$ and the $c_{\lambda }$ correspond to the
fiducial state (\ref{ecd3}). The matrix element in Eq.(\ref{fp}) is

\begin{equation}
\left(\left\langle \xi\right\vert Z_{\gamma}X_{\delta}\left\vert
\xi\right\rangle \right)^{-1}|_{\mathbf{h}\left(\gamma,\delta\right)=\mathbf{%
q}} = 2^{\frac{1}{9}\sum_{k,l}q_{kl}-N} \mathrm{e}^{\frac{i\pi}{9}%
(q_{21}-q_{22}+q_{12}+5q_{11}-9q_{01}-3(q_{10}-q_{02}+q_{20}))},
\end{equation}
where $q_{kl}=h(k\gamma+l\delta)$ and

\begin{equation}
C_{t_{1}t_{2}t_{1}^{\prime}t_{2}^{\prime}}(\mathbf{q})=\sum_{\gamma,\delta,%
\lambda}\omega^{\gamma\lambda}\delta_{\mathbf{h}\left(\gamma,\delta\right),%
\mathbf{q}}\delta_{h\left(\lambda\right),t_{1}+2t_{2}}\delta_{h\left(2%
\lambda\right),2t_{1}+t_{2}}\delta_{h\left(\lambda-\delta\right),t_{1}^{%
\prime}+2t_{2}^{\prime}}\delta_{h\left(2\lambda-2\delta\right),2t_{1}^{%
\prime}+t_{2}^{\prime}}.
\end{equation}

We have numerically found that for qutrits the minimum error $\sqrt{%
\left\langle \left\langle \mathcal{E}_{min}^{2}\right\rangle \right\rangle }$
also behaves according to Eq. (\ref{Error}). In Fig. 4 we plot the constant $%
\lambda$ for pure and mixed qutrit states, $N=1,2,3$.

\begin{figure}[tbp]
\centering \includegraphics[scale=0.45]{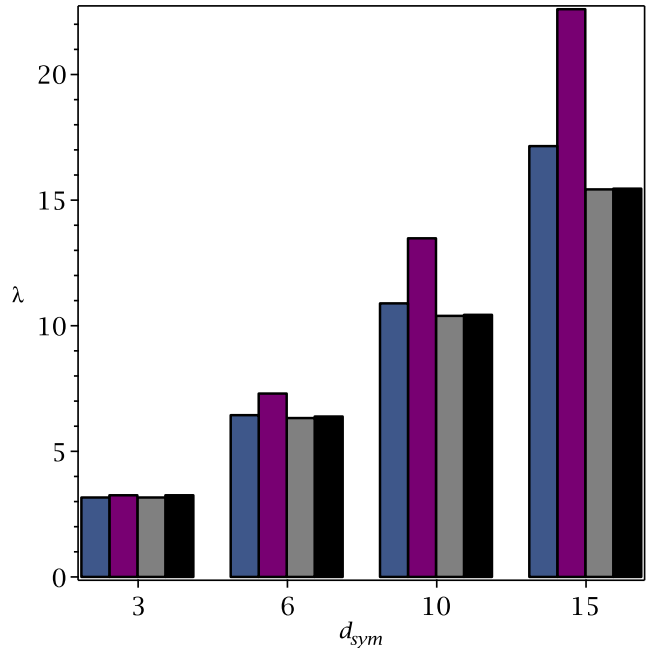}
\caption{ First two columns: the proportionality constant $\protect\lambda$
appearing in the minimum square error (\protect\ref{Error}) for pure (left
column) and mixed (right column) states $d_{sym}=\left(N+1\right)\left(N+2%
\right)/2$ for $N=1,...,4$ qubits corresponding to the reconstruction skill (%
\protect\ref{symrec}); Second two columns: the corresponding values of $%
\protect\lambda$ for SIC tomographic protocol.}
\end{figure}

\subsection{Conclusions}

We have developed a general framework for the analysis of $N$-qudit systems
in the macroscopic limit, which includes:

a) A scheme to organize the information obtained from collective
measurements in form of distributions functions in a discrete
low-dimensional space spanned by the symmetric functions (\ref{hh}). The
analysis of these projected $\tilde{Q}$-functions provides a useful insight
into global properties of multipartite quantum states. It allows, for
instance, to identify the relevant set of collective measurements for a
given state, which are not always obvious from the state decomposition in
the computational basis.

b) A set of collective operators appropriate for characterization of $N$%
-qudit states. The algebraic properties (\ref{Op1})-(\ref{Op2}) of these
operators (\ref{O1}) allow their efficient measurements with further
applications to the collective tomography protocol (\ref{rho1});

c) Explicit expressions for the state reconstruction from collective
measurements, both in the whole $d^{N}$-dimensional space (\ref{rho_rec}), (%
\ref{rho1}) and in the symmetric subspace (\ref{symrec}). The symmetric
tomographic protocol based on projective measurements performed in POVM (\ref%
{Qsym1}) is characterized by a larger MSE than the optimal SIC
reconstruction scheme for N-qubit states. Nevertheless, for larger $d$ -
dimensions the relative difference between both reconstruction methods
becomes smaller. The advantage of the proposed method consists in the
possibility to generate the first rank collective POVM (\ref{Qsym1}) for $N$
qudit systems in a systematic way for an arbitarary number of particles.

A deeper analysis of the projected $N$-qudit $\tilde{Q}$-function and their
applications is in progress and will be published elsewhere.

This work is partially supported by the Grant 254127 of CONACyT (Mexico).

\appendix
we outline the general method for computing $P$-symbols of collective
operators.

Consider a generic (non-Hermitian) collective operator\
\begin{equation}
\sum_{i=1}^{N}Z_{i}^{m}X_{i}^{n}=\hat{s}_{mn},\;m,n\in Z_{d}.  \label{ocscf}
\end{equation}
Any other collective operator can be constructed from (\ref{ocscf}). Taking
into account the explicit form of the kernels (\ref{symbols1})
\begin{eqnarray}
\hat{\Delta}^{(\mp1)}\left(\alpha,\beta\right)=\frac{1}{d^{\frac{1}{2}%
\left(3\pm1\right)N}}\sum_{\gamma,\delta}\omega^{\alpha\delta-\beta\gamma-%
\frac{1}{2}\left(1\pm1\right)\gamma\delta}\left(\left\langle \xi\right\vert
Z_{\mp\gamma}X_{\mp\delta}\left\vert \xi\right\rangle
\right)^{\pm1}Z_{\gamma}X_{\delta},  \label{pmkernel} \\
Tr\hat{\Delta}^{(\mp1)}\left(\alpha,\beta\right)=d^{\frac{1}{2}%
\left(1\mp1\right)N},\;\sum_{\alpha,\beta}\hat{\Delta}^{(\mp1)}\left(\alpha,%
\beta\right)=d^{\frac{1}{2}\left(1\pm1\right)N}\hat{I},
\end{eqnarray}
the $P$ -function of (\ref{ocscf}) is represented as follows according to (%
\ref{symbols1})

\begin{equation}
P_{mn}=\frac{\omega^{mn}}{d^{N}}\left(\left\langle \xi\right\vert
Z^{-m}X^{-n}\left\vert \xi\right\rangle
\right)^{-1}\sum_{i=1}^{N}\omega^{mb_{i}-na_{i}}.  \label{fpoc}
\end{equation}
Observe, that for the fiducial state of the form (\ref{efd})-(\ref{ecsp})
the matrix element $\left\langle \xi\right\vert
Z_{\gamma}X_{\delta}\left\vert \xi\right\rangle $ is symmetric function of
its argument,
\begin{equation}
\left\langle \xi\right\vert Z_{\gamma}X_{\delta}\left\vert \xi\right\rangle
=\left\langle \xi\right\vert Z_{\gamma}X_{\delta}\left\vert \xi\right\rangle
|_{\mathbf{h}\left(\gamma,\delta\right)=\mathbf{p}}.  \label{ZXme}
\end{equation}

The sum $\sum_{i}\omega^{mb_{i}-na_{i}}$ that appear in the above equation
in can be rewritten as

\begin{gather}
\sum_{i=1}^{N}\omega^{mb_{i}-na_{i}}=\sum_{i=1}^{N}\left[\frac{1}{(d-1)!b_{i}%
}+\frac{\omega^{m}}{(d-2)!\left(1-b_{i}\right)}+\frac{\omega^{2m}}{%
-2(d-3)!\left(2-b_{i}\right)}+...\right]  \label{sumom} \\
\left[\frac{1}{(d-1)!a_{i}}+\frac{\omega^{-n}}{(d-2)!\left(1-a_{i}\right)}+%
\frac{\omega^{-2n}}{-2(d-3)!\left(2-a_{i}\right)}+...\right]%
\Pi_{k=0}^{d-1}(k-b_{i})\left(k-a_{i}\right),  \notag
\end{gather}
which is always reduced to the form

\begin{equation}
\sum_{i=1}^{N}a_{i}^{p}b_{i}^{q}=\sum_{k,l=0}^{d-1}B_{k,l}^{\left(p,q%
\right)}h\left(k\alpha+l\beta\right),
\end{equation}
where $B_{k,l}^{\left(p,q\right)}$ are some coefficients. Thus, the $P$%
-function (\ref{fpoc}) is a function of the weights (\ref{hh}).

\textit{Examples}

a) Qubits, $d=2$

The sum (\ref{sumom}) takes the form, $\omega=-1$
\begin{equation}
\sum_{i=1}^{N}\omega^{mb_{i}-na_{i}}=N-2\delta_{m,0}\delta_{n,1}h\left(%
\alpha\right)-4\delta_{m,1}\delta_{n,0}h\left(\beta\right)-4\delta_{m,1}%
\delta_{n,1}h\left(\alpha+\beta\right),
\end{equation}
so that, for the fiducial state (\ref{ecd2})

\begin{equation}
P_{mn}\left(\alpha,\beta\right)=\frac{\sqrt{3}}{2^{N}}\left[%
N-2\left(h\left(\alpha\right)\delta_{m,0}\delta_{n,1}+h\left(\beta\right)%
\delta_{m,1}\delta_{n,0}+h\left(\alpha+\beta\right)\delta_{m,1}\delta_{n,1}%
\right)\right].
\end{equation}
b) Qutrits, $d=3.$

The sum (\ref{sumom}) is significantly more complicated, $\omega=\exp(2\pi
i/3)$,

\begin{gather*}
\sum_{i=1}^{N}\omega^{mb_{i}-na_{i}}=N-\left[\delta_{m,0}\delta_{n,1}e^{-i%
\pi/3}+\delta_{m,0}\delta_{n,2}e^{i\pi/3}\right]h\left(\alpha\right)-\left[%
\delta_{m,0}\delta_{n,1}e^{i\pi/3}+\delta_{m,0}\delta_{n,2}e^{-i\pi/3}\right]%
h\left(2\alpha\right) \\
-\left[\delta_{m,1}\delta_{n,0}e^{i\pi/3}+\delta_{m,2}\delta_{n,0}e^{-i\pi/3}%
\right]h\left(\beta\right)-\left[\delta_{m,1}\delta_{n,0}e^{-i\pi/3}+%
\delta_{m,2}\delta_{n,0}e^{i\pi/3}\right]h\left(2\beta\right) \\
-\left[\delta_{m,1}\delta_{n,2}e^{i\pi/3}+\delta_{m,2}\delta_{n,1}e^{-i\pi/3}%
\right]h\left(\alpha+\beta\right)-\left[\delta_{m,1}\delta_{n,2}e^{-i\pi/3}+%
\delta_{m,2}\delta_{n,1}e^{i\pi/3}\right]h\left(2\alpha+2\beta\right) \\
-\left[-\delta_{m,1}\delta_{n,1}e^{i\pi/3}+\delta_{m,2}\delta_{n,2}e^{-i%
\pi/3}\right]h\left(2\alpha+\beta\right)-\left[\delta_{m,1}\delta_{n,1}e^{-i%
\pi/3}+\delta_{m,2}\delta_{n,2}e^{i\pi/3}\right]h\left(\alpha+2\beta\right).
\end{gather*}

The matrix elements of the operators (\ref{O1}) have the form
\begin{eqnarray}
_{i}\left\langle p\right\vert \hat{O}_{0,l}^{(i)}\left\vert q\right\rangle
_{i}=\delta_{p,q}\left[1-\frac{2}{\left(d-1\right)}\sum_{r=0}^{d-1}\left\{
lr\right\} \left\vert c_{p-r}^{\left(i\right)}\right\vert ^{2}\right],
\label{OpqD} \\
_{i}\left\langle p\right\vert \hat{O}_{\lambda,\lambda m}^{(i)}\left\vert
q\right\rangle _{i}=\delta_{p,q}-\frac{2\tau_{q-p,\lambda}}{d\left(d-1\right)%
}\sum_{r=0}^{d-1}\omega^{-mr\left(q-p\right)}c_{p-r}^{\left(i%
\right)}c_{q-r}^{\left(i\right)\ast},  \label{OpqND}
\end{eqnarray}

where $c_{p}^{\left(i\right)}$ are the expansion coefficients (\ref{ci}),
and $\tau_{q,\lambda}=\sum_{r=0}^{d-1}r\omega^{rq\lambda^{-1}}$, $%
\lambda,m=0,...,d-1$. The trace of two operators (\ref{O1}) has the form

\begin{gather*}
tr\left(\hat{O}_{k,l}\hat{O}_{k^{\prime},l^{\prime}}\right)=N^{2}d^{N}-\frac{%
2N}{\left(d-1\right)d^{N}}\left(\sum_{\alpha,\beta}h\left(k\alpha+l\beta%
\right)+\sum_{\alpha,\beta}h\left(k^{\prime}\alpha+l^{\prime}\beta\right)%
\right) \\
+\frac{4}{\left(d-1\right)^{2}d^{2N}}\frac{1}{d^{N}+1}\left(\sum_{\alpha,%
\beta,\alpha^{\prime},\beta^{\prime}}h\left(k\alpha+l\beta\right)h\left(k^{%
\prime}\alpha^{\prime}+l^{\prime}\beta^{\prime}\right)-\sum_{\alpha,\beta}h%
\left(k\alpha+l\beta\right)h\left(k^{\prime}\alpha+l^{\prime}\beta\right)%
\right) \\
+\frac{4}{\left(d-1\right)^{2}d^{2N}}\sum_{\alpha,\beta}h\left(k\alpha+l%
\beta\right)h\left(k^{\prime}\alpha+l^{\prime}\beta\right),
\end{gather*}
where we have used the SIC POVM condition (\ref{ecsp}).

Taking into account that

\begin{eqnarray}
\sum_{\mu}h\left(\mu\right) & = & \frac{N\left(d-1\right)d^{N}}{2},
\label{B1a} \\
\sum_{\alpha,\beta}h\left(k\alpha+l\beta\right) & = & \frac{%
N\left(d-1\right)d^{2N}}{2}  \label{B1b}
\end{eqnarray}
we obtain

\begin{equation}
tr\left(\hat{O}_{k,l}\hat{O}_{k^{\prime},l^{\prime}}\right)=-\frac{N^{2}d^{N}%
}{d^{N}+1}+\frac{4}{d^{N}\left(d-1\right)^{2}\left(d^{N}+1\right)}%
\sum_{\alpha,\beta}h\left(k\alpha+l\beta\right)h\left(k^{\prime}\alpha+l^{%
\prime}\beta\right).  \label{B2}
\end{equation}

In case when $k\neq \lambda k^{\prime }$ and $l\neq \lambda l^{\prime }$, $%
\lambda =1,...,d-1$ we can use new variables $k\alpha +l\beta =\mu $, $%
\lambda (k^{\prime }\alpha +l^{\prime }\beta )=\nu $ and sum over the new
variables $\mu $, $\nu $ so that the last sum in (\ref{B2}) is just a square
of (\ref{B1a}) and thus

\begin{equation}
tr\left( \hat{O}_{k,l}\hat{O}_{k^{\prime },l^{\prime }}\right) =0;
\end{equation}%
if $k=\lambda k^{\prime }$ and $l=\lambda l^{\prime }$, then

\begin{equation}
tr\left(\hat{O}_{k,l}\hat{O}_{k^{\prime},l^{\prime}}\right)=-\frac{N^{2}d^{N}%
}{d^{N}+1}+\frac{4}{\left(d-1\right)^{2}\left(d^{N}+1\right)}%
\sum_{\mu}h\left(\mu\right)h\left(\lambda\mu\right).
\end{equation}
Taking into account the representation (\ref{fP}) and the form of the
P-function (\ref{PO})

\begin{gather*}
\hat{O}_{k,l}=\sum_{\alpha,\beta}\frac{1}{d^{N}}\left[N-\frac{2}{d-1}%
h\left(k\alpha+l\beta\right)\right]\Delta^{(-1)}\left(\alpha,\beta\right) \\
=N\hat{I}-\frac{2}{d^{2N}\left(d-1\right)}\sum_{\gamma,\delta}\omega^{-%
\gamma\delta}\left[\sum_{\alpha,\beta}h\left(k\alpha+l\beta\right)\omega^{%
\alpha\delta-\beta\gamma}\right]\left\langle \xi\right\vert
Z_{-\gamma}X_{-\delta}\left\vert \xi\right\rangle Z_{\gamma}X_{\delta}
\end{gather*}
we immediately observe that
\begin{equation}
\lbrack\hat{O}_{k,0},\hat{O}_{\lambda k,0}]=[\hat{O}_{0,l},\hat{O}%
_{0,\lambda l}]=0.
\end{equation}
If $k\neq0$, $l\neq0$ for $d>2$ then after a change of variables

\begin{gather*}
\alpha=\left(\mu+\nu\right)\left(2k\right)^{-1}, \\
\beta=\left(\mu-\nu\right)\left(2l\right)^{-1},
\end{gather*}
where $\left(2k\right)^{-1}$ is understood as the inverse element in $Z_{d}$%
, we represent $\hat{O}_{k,l}$ as

\begin{equation}
\hat{O}_{k,l}=N\hat{I}-\frac{2}{d^{N}\left(d-1\right)}\sum_{\gamma}%
\omega^{kl^{-1}\gamma^{2}}\left[\sum_{\mu}h\left(\mu\right)\omega^{-\mu%
\gamma l^{-1}}\right]\left\langle \xi\right\vert
Z_{-\gamma}X_{kl^{-1}\gamma}\left\vert \xi\right\rangle
Z_{\gamma}X_{-k^{-1}\gamma l},
\end{equation}
so that

\begin{equation}
\left[ \hat{O}_{k,l},\hat{O}_{\lambda k,\lambda l}\right] =0.
\end{equation}%
It is worth noting that the average values of the collective operators (\ref%
{O1}) in the states (\ref{absym1}) are proportional to the corresponding $P$%
-symbols (\ref{PO})
\begin{equation}
\left\langle \phi _{\mathbf{h}\left( \alpha ,\beta \right) =\mathbf{m}%
}\right\vert \hat{O}_{k,l}\left\vert \phi _{\mathbf{h}\left( \alpha ,\beta
\right) =\mathbf{m}}\right\rangle =\frac{d^{N}}{d+1}P_{\hat{O}_{k,l}}.
\end{equation}

Here we present explicit expressions for $R_{\mathbf{q}}^{(d)}$-functions
for qubits and qutrits:

\begin{gather}
R_{\mathbf{q}}^{(2)}=\frac{N!}{\left(\frac{q_{10}+q_{01}-q_{11}}{2}%
\right)!\left(\frac{2N-q_{10}-q_{01}-q_{11}}{2}\right)!\left(\frac{%
q_{10}-q_{01}+q_{11}}{2}\right)!\left(\frac{q_{01}+q_{11}-q_{10}}{2}\right)!}%
,  \label{R2} \\
0\leq q_{kl}=h\left(k\gamma+l\delta\right)\leq N,
\end{gather}
and
\begin{gather}
(R_{\mathbf{q}}^{\left(3\right)})^{-1}=\frac{1}{N!}\left(N-\frac{%
q_{10}+q_{20}+q_{01}+q_{02}+q_{11}+q_{22}+q_{21}+q_{12}}{9}\right)!\times
\label{R3} \\
\left(\frac{2q_{10}-q_{20}+2q_{01}-q_{02}-q_{11}+2q_{22}-q_{21}-q_{12}}{9}%
\right)!\times  \notag \\
\left(\frac{2q_{10}-q_{20}-q_{01}+2q_{02}-q_{11}-q_{22}+2q_{21}-q_{12}}{9}%
\right)!\times  \notag \\
\left(\frac{2q_{10}-q_{20}-q_{01}-q_{02}+2q_{11}-q_{22}-q_{21}+2q_{12}}{9}%
\right)!\times  \notag \\
\left(\frac{-q_{10}+2q_{20}+2q_{01}-q_{02}-q_{11}-q_{22}-q_{21}+2q_{12}}{9}%
\right)!\times  \notag \\
\left(\frac{-q_{10}+2q_{20}-q_{01}+2q_{02}+2q_{11}-q_{22}-q_{21}-q_{12}}{9}%
\right)!\times  \notag \\
\left(\frac{-q_{10}+2q_{20}-q_{01}-q_{02}-q_{11}+2q_{22}+2q_{21}-q_{12}}{9}%
\right)!\times  \notag \\
\left(\frac{-q_{10}-q_{20}+2q_{01}-q_{02}+2q_{11}-q_{22}+2q_{21}-q_{12}}{9}%
\right)!\times  \notag \\
\left(\frac{-q_{10}-q_{20}-q_{01}+2q_{02}-q_{11}+2q_{22}-q_{21}+2q_{12}}{9}%
\right)!  \notag \\
0\leq q_{kl}=h\left(k\gamma+l\delta\right)\leq N,
\end{gather}

In the limit $N\gg1$ the function $R_{\mathbf{q}}^{\left(d\right)}$ tends to
a Gaussian form localized in the vicinity of the point $\mathbf{q}_{0}=\frac{%
d-1}{2}(N,...,N)$, for instance,

\begin{gather}
R_{q_{10}q_{01}q_{11}}^{(2)}\sim\exp(-2(\mathbf{q}-\mathbf{q}_{0})^{2}/N), \\
R_{\mathbf{q}}^{\left(3\right)}\sim\exp(-\sum_{i=0}^{1}\sum_{j=0}^{i+1}\frac{%
\left[(q_{ij}-N)^{2}+(q_{2i2j}-N)^{2}-(q_{ij}-N)(q_{2i2j}-N)\right]}{N}).
\end{gather}

Any product of collective operators (\ref{O1}) can be directly expanded in
the monomial basis (\ref{ZX}):
\begin{equation}
\prod_{k,l}O_{k,l}^{\mathtt{p}_{k,l}}=\sum_{\gamma,\delta}b_{\mathbf{h}%
\left(\gamma,\delta\right),\mathtt{p}}Z_{\gamma}X_{\delta}.  \label{D1}
\end{equation}
The required representation of the symmetrized monomials in terms of
measurables operators (\ref{O1})

\begin{equation}
\sum_{\mu,\nu}Z_{\mu}X_{\nu}\delta_{\mathbf{h}\left(\mu,\nu\right),\mathbf{m}%
}=\sum_{\mathbf{\mathtt{p}}}C_{\mathbf{\mathtt{p}}}^{\left(\mathbf{m}%
\right)}\prod_{k,l}O_{k,l}^{\mathtt{p}_{k,l}},\quad\sum_{k,l}\mathtt{p}%
_{k,l}\leq\sum_{k,l}m_{kl},\;  \label{D2}
\end{equation}
is thus obtained by substituting (\ref{D1}) into (\ref{D2}) and comparing
the coefficients of the same basis elements. This leads to the following set
of equations for the coefficients $C_{\mathbf{\mathtt{p}}}^{\left(\mathbf{m}%
\right)}$

\begin{equation}
\delta_{\mathbf{h}\left(\mu,\nu\right),\mathbf{m}}=\sum_{\mathbf{\mathtt{p}}%
}C_{\mathbf{\mathtt{p}}}^{\left(\mathbf{m}\right)}b_{\mathbf{h}%
\left(\mu,\nu\right),\mathtt{p}}.  \label{D3}
\end{equation}
In case of qubits, (\ref{D1}) adquires the following form for diagonal
operators
\begin{equation}
\sum_{\mu}Z_{\mu}\delta_{h\left(\mu\right),k}=\sum_{p=0}^{k}C_{p}^{\left(k%
\right)}S_{z}^{p}.
\end{equation}
Taking into account that the collective spin operator in the computational
basis has the form

\begin{equation}
S_{z}=\sum_{\nu}\left(N-h\left(\nu\right)\right)\left\vert \nu\right\rangle
\left\langle \nu\right\vert ,
\end{equation}
we arrive at the following expression for the coefficient $b$

\begin{equation}
b_{h\left(\mu\right),p}=\frac{1}{2^{N}}\sum_{\nu}\left(N-h\left(\nu\right)%
\right)^{p}\chi\left(\mu\nu\right)=\frac{1}{2^{N}}\sum_{m=0}^{N}{\binom{N}{m}%
}\left(N-m\right)^{p}g^{\left(2\right)}(m),
\end{equation}
where
\begin{equation}
g^{(2)}(m)=\sum_{\nu}\left(-1\right)^{\mu\nu}\delta_{h\left(\nu\right),m},
\end{equation}
is a discrete special function, which can be expressed in terms of $%
P_{n}^{\left(\alpha,\beta\right)}\left(z\right)$ Jacobi polynomials. The
system (\ref{D3}) takes the form

\begin{equation}
\delta_{h\left(\mu\right),m}=\sum_{p\leq
m}C_{p}^{\left(k\right)}b_{h\left(\mu\right),p},
\end{equation}
leading in particular to

\begin{eqnarray}
\sum_{\mu}Z_{\mu}\delta_{h\left(\mu\right),1} & = & \hat{S}_{z}, \\
\sum_{\mu}Z_{\mu}\delta_{h\left(\mu\right),2} & = & \hat{S}_{z}^{2}-N\hat{I}.
\end{eqnarray}
Similar calculations can be carried out for qutrits:

\begin{equation}
\sum_{\mu}Z_{\mu}\delta_{h\left(\mu\right),m_{1}}\delta_{h\left(2\mu%
\right),m_{2}}=\sum_{p_{1}+p_{2}\leq
m_{1}+m_{2}}C_{p_{1},p_{2}}^{\left(m_{1},m_{2}\right)}\hat{O}_{0,1}^{p_{1}}%
\hat{O}_{0,2}^{p_{2}},
\end{equation}
where

\begin{equation}
\hat{O}_{0,l}^{p}=\sum_{\nu}\left(N-\sum_{\varepsilon}h\left(l\varepsilon%
\right)\left\vert c_{\nu-\varepsilon}\right\vert ^{2}\right)^{p}\left\vert
\nu\right\rangle \left\langle \nu\right\vert ,
\end{equation}
being $c_{\nu}$ expansion coefficients (\ref{ciN}) of the fiducial state in
the computational basis.

Thus the coefficients $b$ have the form

\begin{eqnarray}
b_{\mathbf{h}\left(\mu\right),p_{1},p_{2}} & = & \frac{1}{%
3^{N}2^{p_{1}+p_{2}}}\sum_{m_{1},m_{2}}\left(m_{2}-m_{1}\right)^{p_{1}}%
\left(N-m_{1}\right)^{p_{2}}g^{(3)}(m_{1},m_{2}) \\
g^{(3)}(m_{1},m_{2}) & = &
\sum_{\nu}\delta_{h\left(\nu\right),m_{1}}\delta_{h\left(2\nu\right),m_{2}}%
\omega^{-\mu\nu},\;\omega=\exp(2\pi i/3),
\end{eqnarray}
where we used the explicit expansion (\ref{ecd3}). The set of equations (\ref%
{D3}) to be inverted is now
\begin{equation}
\delta_{\mathbf{h}\left(\mu\right),\mathbf{m}}=\sum_{p_{1}+p_{2}\leq
m_{1}+m_{2}}C_{p_{1},p_{2}}^{\left(m_{1},m_{2}\right)}b_{\mathbf{h}%
\left(\mu\right),p_{1},p_{2}}.
\end{equation}
In particular, we find,

\begin{eqnarray}
\sum_{\mu}Z_{\mu}\delta_{h\left(\mu\right),1} & = & 2\omega\hat{O}_{0,1}+2%
\hat{O}_{0,2},  \label{Z3a} \\
\sum_{\mu}Z_{\mu}\delta_{h\left(\mu\right),2} & = & 4\left(\omega\hat{O}%
_{0,1}+\hat{O}_{0,2}\right)^{2}+2\left(\omega^{2}\hat{O}_{0,1}+\hat{O}%
_{0,2}\right).  \label{Z3b}
\end{eqnarray}

The performance of a reconstruction scheme can be measured by the
statistical average of the Hilbert-Schmidt distance between the real $%
\rho_{s}$ and estimated $\tilde{\rho}_{s}$ states,
\begin{equation}
\langle\mathcal{E}^{2}\rangle=\langle Tr[(\rho_{s}-\tilde{\rho}%
_{s})^{2}]\rangle,
\end{equation}
when only a finite number of copies ($M$) are involved in the measurement
process \cite{natcom}.

The multinomial measurement statistics associated to the protocol (\ref%
{symrec})
\begin{equation}
\mathcal{P}\propto\prod_{\mathbf{p}}\tilde{\sigma}_{\mathbf{p}}^{n_{\mathbf{p%
}}},\quad\sum_{\mathbf{p}}n_{\mathbf{p}}=M,
\end{equation}
allows to relate the estimated probabilities $\tilde{\sigma}_{\mathbf{p}}$
with the corresponding frequencies $n_{\mathbf{p}}/M$, according to
\begin{equation}
\langle n_{\mathbf{p}}\rangle=M\tilde{\sigma}_{\mathbf{p}},~\langle n_{%
\mathbf{p}}^{2}\rangle=M\tilde{\sigma}_{\mathbf{p}}(1+(M-1)\tilde{\sigma}_{%
\mathbf{p}}),~\langle n_{\mathbf{p}}n_{\mathbf{q}}\rangle=M(M-1)\tilde{\sigma%
}_{\mathbf{p}}\tilde{\sigma}_{\mathbf{q}}.
\end{equation}
Using the explicit reconstruction form (\ref{symrec}) we obtain the square
error in terms of the deviation between probabilities $\sigma_{\mathbf{p}}$
and their estimates $\tilde{\sigma}_{\mathbf{p}}$,
\begin{equation}
\left\langle \mathcal{E}^{2}\right\rangle =\sum_{\mathbf{p},\mathbf{q}}%
\mathcal{A}_{\mathbf{p},\mathbf{q}}\Delta\sigma_{\mathbf{p}}\Delta\sigma_{%
\mathbf{q}},
\end{equation}
where $\Delta\sigma_{\mathbf{p}}=\sigma_{\mathbf{p}}-\tilde{\sigma}_{\mathbf{%
p}}$ and
\begin{equation}
\mathcal{A}_{\mathbf{p},\mathbf{q}}=d^{2N}\left(R_{\mathbf{p}}^{(d)}R_{%
\mathbf{q}}^{(d)}\right)^{-1}Tr\left(\hat{\Delta}_{s}^{(1)}\left(\mathbf{p}%
\right)\hat{\Delta}_{s}^{(1)}\left(\mathbf{q}\right)\right).
\end{equation}
Taking into account the redundancy (\ref{redundance}) and normalization (\ref%
{psym}) conditions we obtain the square error in terms of the deviation
between the independent probabilities $\sigma_{\mathbf{p}}^{\prime}$ and
their estimates $\tilde{\sigma}_{\mathbf{p}}^{\prime}$
\begin{equation}
\left\langle \mathcal{E}^{2}\right\rangle =\sum_{\mathbf{p^{\prime}},\mathbf{%
q^{\prime}}}\mathcal{A}_{\mathbf{p^{\prime}}\mathbf{q^{\prime}}%
}^{\prime}\Delta\sigma_{\mathbf{p^{\prime}}}\Delta\sigma_{\mathbf{q^{\prime}}%
},  \label{errorq}
\end{equation}
where the sum is taken only on the indexes $\mathbf{p^{\prime},q^{\prime}}$
labeling the independent probabilities; the coefficients $\mathcal{A}_{%
\mathbf{p^{\prime}}\mathbf{q^{\prime}}}^{\prime}$ are not explicitly given
here due to of their cumbersome form.

The minimum square error (MSE) is given by the Cramér-Rao bound \cite%
{Helstrom},
\begin{equation}
\left\langle \mathcal{E}^{2}\right\rangle \geq Tr(\mathcal{A}^{\prime}%
\mathcal{F}^{-1}),
\end{equation}
were $\mathcal{A}^{\prime}$ is $\left(d_{sym}^{2}-1\right)\times%
\left(d_{sym}^{2}-1\right)$ matrix with the elements $\mathcal{A}_{\mathbf{%
p^{\prime}}\mathbf{q^{\prime}}}^{\prime}$, while $\mathcal{F}$ is the Fisher
matrix for the independent probabilities,
\begin{equation}
\mathcal{F}_{\mathbf{p^{\prime}}\mathbf{q^{\prime}}}=\left\langle \frac{%
\partial\ln\mathcal{P}}{\partial\sigma_{\mathbf{p^{\prime}}}^{\prime}}\frac{%
\partial\ln\mathcal{P}}{\partial\sigma_{\mathbf{q^{\prime}}}^{\prime}}%
\right\rangle .
\end{equation}
To estimate the average minimum square error we numerically compute the
statistical mean value over the symmetric states,
\begin{equation}
\sqrt{\left\langle \left\langle \mathcal{E}_{min}^{2}\right\rangle
\right\rangle }=\sqrt{\left\langle Tr(\mathcal{A}^{\prime}\mathcal{F}%
^{-1})\right\rangle }.  \label{mse}
\end{equation}
.

\end{document}